\documentclass[aps,prl,twocolumn,groupedaddress,amsmath,amssymb]{revtex4-1}

\usepackage{amsmath}
\usepackage{graphicx}
\usepackage{amsfonts}
\usepackage{amssymb}
\usepackage{color}
\usepackage{dcolumn}
\usepackage{bm}

\newcommand{\beq}{\begin{equation}}
\newcommand{\eeq}{\end{equation}}
\newcommand{\beqa}{\begin{eqnarray}}
\newcommand{\eeqa}{\end{eqnarray}}

\begin{document}


\title{Confinement as a tool to probe amorphous order} 

\author{C. Cammarota}
\altaffiliation[Also at ]{LPTMC, Tour 12-13/13-23, Bote 121, 4, Place Jussieu, 75252 Paris Cedex 05, France}
\author{G. Gradenigo}%
\author{G. Biroli}
\affiliation{%
IPhT, CEA/DSM-CNRS/URA 2306, CEA Saclay, F-91191 Gif-sur-Yvette Cedex, France
}%

%

\date{\today}

\begin{abstract}
We study the effect of confinement on glassy liquids using Random First Order Transition theory as framework. We show that the characteristic length-scale above which confinement effects become negligible is related to the point-to-set length-scale introduced to measure the spatial extent of amorphous order in super-cooled liquids. By confining below this characteristic size, the system becomes a glass. Eventually, for very small sizes, the effect of the boundary is so strong that any collective glassy behavior is wiped out. We clarify similarities and differences between the physical behaviors induced by confinement and by pinning particles outside a spherical cavity (the protocol introduced to measure the point-to-set length). Finally, we discuss possible numerical and experimental tests of our predictions.
\end{abstract}

\pacs{Valid PACS appear here}
\maketitle

The search for a growing static length accompanying the slowing down of the dynamics of super-cooled liquids is a {\it leit-motif} and a key open issue in the study of the glass transition. 
The super-Arrhenius behavior of the relaxation time is indeed a hint that such a length exists: growing energy barrier should be related to an increasing cooperativity and, hence, to a growing static length as proposed already long-time ago by Adam and Gibbs \cite{adamgibbs}. 
Recently, this intuition was put on a rigorous basis by Montanari and Semerjian \cite{montanarisemerjian}. 
Their result was obtained using the cooperative length-scale, $\ell_{PS}$, that measures the spatial extent of amorphous order and that was originally introduced in~\cite{BouBir04} to characterize the spatial structure of the so-called mosaic state envisioned for super-cooled liquids by the Random First Order Transition (RFOT) theory \cite{KTW}. The definition of $\ell_{PS}$, called point-to-set (PS) length, is the following: take a typical equilibrium configuration, freeze the positions of all particles outside a sphere centered around a given point and study how the thermodynamics of the remaining particles, inside the sphere, is influenced by this amorphous boundary condition \cite{BouBir04,mezardmontanari}; $\ell_{PS}$ is the smallest radius 
of the sphere at which the boundary has no longer any effect on the configuration at the center. 
As its definition above makes clear, $\ell_{PS}$ is quite difficult to measure. It can be obtained by numerical simulations, but for rather high temperatures only \cite{CGV,BBCGV,SauTar10,BeKo_PS,HocRei12}, because of equilibration problems \cite{BeKo_PS,CGVdyn}. Therefore, one can only access its first increase in a regime where 
it should not play an important role in determining the dynamics; only in the--deeply supercooled--activated regime $\ell_{PS}$ should be directly linked to the growth of the relaxation time \cite{LubWol_rev,BoBi_rev}. The way out of this {\it impasse} would be measuring such a length in experiments on molecular liquids close to the glass transition. However, this is extremely challenging; no experimental apparatus for doing that has been devised so far. It could be done in colloids; although this is interesting {\it per se}, the range of available time-scales would remain restricted to the first 6-8 decades of slowing down of the dynamics, as in simulations. \\
Actually, there might be an alternative and simpler way to measure a growing static length in super-cooled liquids. In the last twenty years, a large experimental effort crystallized in the
study of the role of spatial confinement on glassy dynamics. Indeed, if 
the glass transition is related to the growth of a static length, the study of confined liquids may unveil the existence of such a length by measuring the smallest confinement linear size, $\ell_{C}$, such that bulk behavior is recovered. 
The idea, as for finite size scaling in critical phenomena, is to use the possibility of varying the system size as an investigation tool. Unfortunately several difficulties get along the way. In particular the interaction between the boundary and the confined fluid and the possible change of density inside the confining region lead to non-universal behavior even for the simpler case of the
melting-freezing transition \cite{alba}. In the case of confined super-cooled liquids 
the glass transition temperature has been found to either increase of decrease as a function of the confinement length scale depending on the experimental system \cite{alba,mckenna}; no clear indication of a growing static length could be found. 
It was not understood, however, whether this is due to an intrinsic inability of $\ell_{C}$ or just to the practical complications cited above. Results obtained in numerical simulations and for colloidal systems point toward the latter possibility \cite{confcoll,kobconf}.
Theoretically, the distinction between $\ell_{C}$ and $\ell_{PS}$ is subtle and boils down to the difference in the boundary conditions used to study the behavior of a confined fluid. For $\ell_{C}$, the boundary (henceforth called random and denoted RB) is essentially formed by a rough wall that bears no correlation besides the short-range ones due to steric constraint. For $\ell_{PS}$, instead, the boundary (henceforth called amorphous and denoted AB) is obtained by freezing particles from an equilibrium configuration at temperature $T$---the hunch is that this protocol quenches very subtle correlations and, hence, the pinned particles at the boundary act as a pinning field that forces the configuration inside the cavity to be a in a given amorphous state for $\ell<\ell_{PS}$. 
In this work we clarify similarities and differences in the physical behaviors of confined liquids with random and amorphous boundary conditions using RFOT theory as a framework \cite{LubWol_rev,BoBi_rev}. 
We have found that $\ell_{C}$ and $\ell_{PS}$ increase in a similar fashion (the former being smaller than the latter) but that the corresponding confined systems respectively behave very differently
 below $\ell_{C}$ and $\ell_{PS}$.
Our results, which are also relevant for recent studies on pinning particles from equilibrium and from random configurations \cite{BeKo_PS,CB_pinning,ProKar11,kunipinning,krakoviackpinning, ParKar12,szamelpinning,charbonneautarjus2,panino,parisi,sandalo}, demonstrate that confinement is indeed a way to probe 
the length-scale associated with the spatial extent of amorphous order in super-cooled liquids.\\
Let us start with some heuristic arguments that will be backed later by analytical computations. 
RFOT theory explains the static and dynamic properties of supercooled liquids in terms of the
competition between the huge number of possible amorphous states in which a 
liquid can freeze, measured by the configurational entropy density $s_c(T)$, and the tendency to sample states with low free energy \cite{LubWol_rev,BoBi_rev}. In order to measure $\ell_{PS}$, one pins all particles outside a spherical cavity; by doing that the number of possible states in which the particles inside the cavity can arrange 
is diminished and, correspondingly, the total configurational entropy inside the cavity decreases and reads at leading order in $\ell$:  
$s_c \frac{4\pi}{3}\ell^3 -4\pi Y_{PS}\ell^{\theta_{PS}}$ \cite{SFGS}. 
The last term is a surface contribution, hence $\theta_{PS} \le 2$ (the $4\pi$ is included in reference to the simplest case $\theta_{PS}=2$); it is thought to originate from the boundary free-energy mismatch between the subset of states which are incompatible at the boundary with the initial configuration used to pin particles. By decreasing $\ell$, fewer and fewer states remain compatible. For $\ell<\ell_{PS}$ only 
the one corresponding to the initial configuration survives. The point-to-set length is therefore directly related to the configurational entropy and reads $\ell_{PS}=(3Y_{PS}/s_c)^{1/(3-\theta_{PS})}$. It represents, within RFOT, the typical linear size over which the system is amorphously ordered---"the mosaic's tile length". The confinement set-up is very similar to the previous one but with the
crucial difference that the boundary is featureless, {\it i.e.} it equally disfavors all states. 
The total configurational entropy is expected to decrease also in this case  as $s_c \frac{4\pi}{3}\ell^3 -4 \pi Y_C\ell^{\theta_C}$ (as found for the one dimensional Kac Random Energy Model in \cite{FPRT}). It is reasonable to assume, and it is in agreement with 
our analytical findings, that $\theta_C=\theta_{PS}$ and $Y_C\lesssim Y_{PS}$, {\it i.e.} confinement leads to a decrease of configurational entropy similar to the AB case. 
We define the confinement length as the value of $\ell$ at which the configurational entropy inside the cavity vanishes, which leads to the result:  $\ell_C=(3 Y_{C}/s_c)^{1/(3-\theta_C)}$.
In the AB case, and for $\ell<\ell_{PS}$, the system is frozen in the only state compatible with the boundary condition; whereas instead in the RB case, for $\ell<\ell_{C}$, it can sample all the lowest free energy states available, whose free energy difference is $O(1)$. This regime is exactly the analog of the one expected below $T_K$, the so-called one step replica symmetry breaking phase. 
Thus, decreasing the confinement length is tantamount to lowering the temperature for bulk systems with the important difference that, since the system is finite, only a cross-over and not a true phase transition happens at $\ell=\ell_{C}$. This analogy suggests that a mode-coupling cross-over should also take place at a length $\ell_{MCT}$ ($>\ell_{C}$), as indeed found in microscopic computations in \cite{krakoviackpinning}.
\begin{figure}
\hspace{-0.3cm}
\includegraphics[scale=0.35,angle=-90]{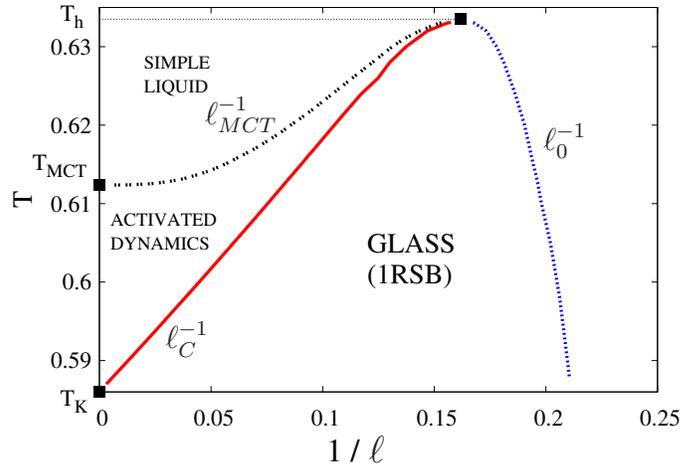}
\put(-118,-178){\Large $\ell$}
\put(-52,-60){\large $\ell^{-1}_0$}
\put(-180,-50){\large $\ell^{-1}_{MCT}$}
\put(-195,-135){\large $\ell^{-1}_{C}$}
\caption{\label{fig3} Confinement "Phase diagram": the continuous line (red) denotes a finite size glass transition at $\ell_C$, the dashed line on the left (black) is the MCT cross-over at $\ell_{MCT}$, and the dotted one on the right (blue) indicates a continuous glass transition at $\ell_0$.}
\vspace{-0.7cm}
\end{figure}
Our previous arguments suggest that although $\ell_{C}<\ell_{PS}$, these length-scales are proportional to each other and increase similarly when the configurational entropy diminishes: if $s_c(T)\propto T-T_K$ for $T\rightarrow T_K$ then they both diverge as $1/(T-T_K)^{1/(3-\theta)}$. We expect that by decreasing 
$\ell$ the relaxation time increases, first because of Mode-Coupling effects and then because of the RFOT-Adam-Gibbs mechanism \cite{LubWol_rev,BoBi_rev} that relates the 
decrease of $s_c$ to the increase of the relaxation time.
When $\ell$ becomes of the order of $\ell_{PS}$ or $\ell_{C}$, in the AB and RB cases respectively, the relaxation timescale should start decreasing---faster in the 
AB case because the system has just to sample one given state \cite{CGVdyn}, 
slower in the RB case where collective rearrangements, corresponding to inter-state dynamics, still go on but involve a smaller number of particles (see EPAPS and  \cite{footnote}). Collective glassy behavior is expected to disappear at very high temperature or very small $\ell$. A sketch of the resulting phase diagram, which actually correspond to the analytical solution we shall present later,  is shown in Fig.1. \\
We now present our analytical investigation of confinement with RB conditions; the AB case was treated in \cite{franzmontanari}. Our starting point is the replica free energy functional $F[q_{ab}(x)]$ already used several times to analyze the glass transition \cite{LubWol_rev}; the spatially varying field $q_{ab}(x)$ is defined for $a<b$, where $a$ and $b$ denote the replica indices running from $1$ to $n$, where $n\rightarrow0$. Note that the random boundary condition acts as an external quenched disorder, this is why one ends 
up with $n\rightarrow 0$ replicas even for a system that does not contain any quenched disorder.
\begin{figure}
\includegraphics[scale=0.35,angle=-90]{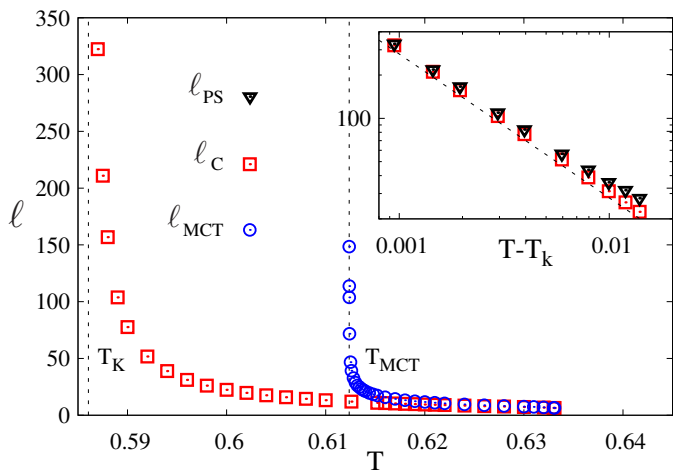}
\put(-192,-35){\large $\ell$}
\put(-190,-61){\large $\ell$}
\put(-200,-85){\large $\ell$}
\put(-260,-85){\Large $\ell$}
\caption{\label{fig1} The lengths $\ell_C$ (squares) and $\ell_{MCT}$ (circles) are plotted as a function of $T$ in the case of random boundaries. The divergence of $\ell_{MCT}$ is proportional to $1/(T-T_{MCT})^{1/4}$ (not shown).
Inset: the behaviour of $\ell_C$ (squares) as a function of $T-T_K$ is compared to that of $\ell_{PS}$ (triangles): the straight line indicates the common power law $1/T-T_K$.}
\vspace{-0.7cm}
\end{figure}
In previous analyses, two forms have been used: a Ginzburg-Landau one \cite{DzSmWo}
and another obtained by analysis based on the Kac-limit \cite{franz}. We focus on the latter because it has the advantage of corresponding to a well-defined model---a disordered p-spin Kac system (see \cite{franz} and the EPAPS for details and \cite{footnote}). Since the Ginzburg-Landau action can be recovered making a gradient and a field expansion, our results are not restricted to this specific choice of $F[q_{ab}(x)]$.
For the p-spin Kac model  the random boundary condition can be explicitly taken into account by requiring that all the 
spins outside the cavity are equal to a random configuration, {\it i.e.} sampled from the infinite temperature Boltzmann measure. One can also show that taking instead a configuration with e.g. all spins up is statistically equivalent. This is natural because from the point of view of an amorphous state a random boundary condition or a non-disordered one are statistically equal. The analog for particle systems of this result is that random boundary conditions (obtained from high $T$ configurations) or rough walls should all be equivalent as far as collective glassy effects are concerned. From the replica point of view, the random boundary conditions lead to the constraint $q_{ab}(x)=1$ $\forall a,b$ outside the cavity.
 As in previous studies, we focus on two {\it ans\"atze} for the form of  $q_{ab}(x)$: one is replica symmetric (RS) $q_{ab}(x)=q_0$ $\forall a,b$ and the other is one step replica symmetry breaking (1RSB), {\it i.e.} replica are collected in $n/m$ groups and $q_{ab}(x)=q_1(x)$ for replica inside the same group and $q_{ab}(x)=q_0(x)$ otherwise. The physical meaning of these solutions are the usual ones: when only the RS solution is present the liquid is simple and not glassy, when the 1RSB solution at $m=1$ appears an exponential number of metastable states emerges (this is related to the Mode-Coupling transition). 
From the derivative of $F[q_{ab}(x)]$ in $m=1$ \cite{remi} one can obtain the configurational entropy which vanishes  when the 1RSB solution with $m<1$ starts to extremize $F[q_{ab}(x)]$, {\it i.e} the system is in the glass phase. 
The form of  $F[q_{ab}(x)]$ within the 1RSB ansatz (the RS can be recovered imposing $q_0=q_1$) reads $F[q_{ab}(x)]=\int d^3x \mathcal{L}_{_{1\textrm{RSB}}}(q_0(x),q_1(x),m)$ where 
\begin{widetext}
\begin{equation}
\mathcal{L}_{_{1\textrm{RSB}}}=(1-m)\frac{\beta^2}{2} f(q_1\ast\psi) + m \frac{\beta^2}{2} f(q_0\ast\psi) - \frac{1}{2} \frac{q_0}{1-(1-m)q_1-m q_0} -\frac{m-1}{2m}\log(1-q_1)-\frac{1}{2m}\log[1-(1-m)q_1-mq_0], \nonumber
\label{lgn-1RSB-random}
\end{equation}
\end{widetext}
and $\psi(x)$ is a normalized three-dimensional Gaussian and $f(q)$ is a function defined
as $f(q)\equiv \frac{q^p}{2}$ with $p=3$ (we considered the $p=3$ Kac-spin model).
The notation $q\ast\psi(x)$ indicates the convolution $\int d^3y \psi(y-x) q(x)$. 
By extremizing $F[q_{ab}(x)]$ with respect to $q_0,q_1$ and $m$, assuming spherical symmetry around the origin and using the boundary conditions $q_0=q_1=1$ outside the cavity
one obtains the inhomogeneous equations determining the RS and 1RSB solutions (more details in Ref. \onlinecite{franzmontanari} and the EPAPS). We now present the results that lead to the phase diagram reported in Fig.1. 
At very high temperature we only find the RS solution for any value of $\ell$, {\it i.e} confinement does not induce any glassy behavior. Below a certain temperature, denoted $T_h$ in Fig.1, and above $T_{MCT}$, we find that the cavity radius $\ell$ plays a role similar to the temperature. By decreasing $\ell$ first the system undergoes a MCT transition at $\ell=\ell_{MCT}$, as also found in \cite{krakoviackpinning}, and then the configurational entropy vanishes at $\ell=\ell_{C}$. 
In Fig.1 we have called "simple liquid" the region where only the RS solution is present, 
"activated dynamics" the one characterized by a finite configurational entropy, where the dynamics is activated and (within RFOT) follows an Adam-Gibbs law, and "glass" the one where replica symmetry is broken and the ideal glass phase sets in.    
Fig. 1 shows that the glass transition line becomes continuous and bends downwards. Thus, by confining the system below $\ell_{C}$, the system eventually exits from the glassy phase for a radius equal to $\ell_0$. In this regime the effect of the boundary is overwhelming and destroys the non-trivial free energy landscape. At $\ell=\ell_0$ $q_0$ and $q_1$ approach one another continuously while $m$ remains less than one.
\begin{figure}
\hspace{-0.8cm}
\includegraphics[scale=0.35,angle=-90]{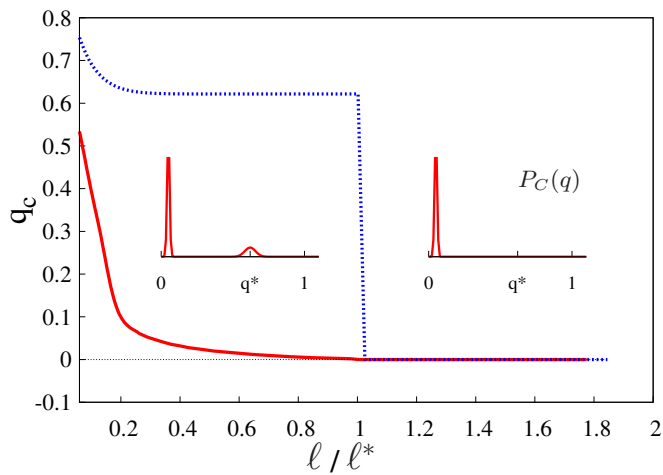}
\put(-140,-177){\Large $\ell$}
\put(-125,-177){\Large $\ell^*$}
\put(-60,-70){$P_C(q)$}
\caption{\label{fig2} Main: Average overlap at the center of the cavity $q_c(\ell/\ell^*)$ as a function of the rescaled cavity radius $\ell/\ell^*$ for random boundary (continuous line), $\ell^*=\ell_c$, and amorphous boundary (dashed line), $\ell^*=\ell_{PS}$ (in this case $T=0.594$, $\ell_c=38.85$, $\ell_{PS}=43.9$). The difference in the confined systems overlap distribution $P_C(q)$ between 
the regime $\ell<\ell_C$ and $\ell>\ell_C$ is shown pictorially.}
\vspace{-0.7cm}
\end{figure}
These results provide a microscopic derivation of the heuristic arguments put forward previously and allow us to determine $\theta_C=2$ and $Y_C$, which is temperature dependent: it is of the order $0.1$ close to $T_K$, and smaller than $Y_{PS}$ of approximatively $10\%$. In Fig. 2 we report the behaviors of $\ell_{C}$ and $\ell_{MCT}$, that look very similar even quantitatively to the analogous ones obtained in the AB case \cite{franzmontanari}. A direct comparison of $\ell_C$ and $\ell_{PS}$ is presented in the inset of Fig.2; they both diverges as a power law $1/(T-T_K)$, but $\ell_{C}$ is slightly smaller than $\ell_{PS}$.\\
The other quantity of interest is the behavior of the average overlap, $\langle q^{(cen)} \rangle$, between two independent equilibrium configurations (in the presence of the same boundary) at the center of the cavity. 
In the AB case, the average overlap jumps discontinuously from a low to a high value at $\ell_{PS}$, whereas in the RB case it starts to increase in a continuous way (with a discontinuous derivative) at $\ell_{C}$, see Fig.3. 
Physically, this is due to the nature of the 1RSB phase and to the probabilistic meaning of its parameter $m$: in the ideal glass (1RSB) phase two equilibrium configurations belong to different states (and have overlap $q_0$) with probability $m$ and belong to
the same state (and have an overlap $q_1$) with probability $1-m$, contrary to the AB case where as soon as the configurational entropy vanishes only one stable configuration is left. 
Since $m\rightarrow 1$ for $\ell \uparrow \ell_C$ the average overlap at $\ell_C$ joins smoothly the one corresponding to the regime $\ell>\ell_C$, where two configurations
are in different states with probability one.  
Since the curve $\langle q^{(cen)}\rangle (R)$ is smooth and does not follow 
a scaling function $f(R/\ell_{C})$ contrary to the AB case
 (it goes as $f(R/\ell_{C})/R$ for $\ell<\ell_C$ and is exponentially small in $\ell$ for $\ell>\ell_C$, see EPAPS), $\langle q^{(cen)}\rangle (R)$ is not suitable to determine $\ell_C$ numerically. 
A better observable is instead the probability distribution of the overlap, which should show 
for $\ell<\ell_C$ two peaks, one at a value $q^{(cen)}_0$ with weight 
$m$ and one at a value $q^{(cen)} _1$ with weight $1-m$, and for $\ell>\ell_C$ only one peak centered in $q^{(cen)}_0$ (fluctuations of $Y_C$ \cite{BBCGV} are expected to make 
the cross-over between these two regimes smooth in real systems).  
In experiments, the easiest protocol to study the effect of confinement consists in measuring the relaxation time, that should first increase substantially approaching $\ell_C$, since the system undergoes a "finite-size glass transition", and then decrease (see the EPAPS for a more detailed discussion).\\
An interesting question, relevant for numerical simulations, is whether periodic boundary conditions are more AB or RB like. Although this deserves further scrutiny, 
a reasonable working hypothesis is that they resemble more to the latter since they do not favor any particular state. Recent numerical simulations have indeed found a non-monotonous dependence of the relaxation time on the system size \cite{GiulioLudo} and an Adam-Gibbs relation between relaxation time and the size dependent configurational entropy \cite{DKS} for super-cooled liquids with periodic boundary conditions.\\
This work, based on RFOT theory, show that "simple" confinement allows to probe the length associated to amorphous order in super-cooled liquids.
We found that the best observables to extract the confinement length are the overlap distribution (which can be likely measured only in simulations) and the relaxation time. We clarified similarities and differences with the case of amorphous boundary conditions, which are actually analogous to the ones found for particles pinned at random from equilibrium and random configurations (see in particularity the similarity between phase diagrams \cite{CB_pinning}).
The conclusion of our work is that, despite the complications faced in the past, 
confinement studies are a route worth pursuing further since they provide a direct access to the length associated to the spatial extent of amorphous order in super-cooled liquids. 
  

\begin{acknowledgments}
We acknowledge support from the ERC grant NPRGGLASS. 
We wish to thank G. Tarjus for helpful discussions. 
\end{acknowledgments}


\begin{thebibliography}{61}
\bibitem{footnote}
{In a particle systems, steric effects induced by the boundary can lead to an enhancement of the microscopic time-scale which could therefore mask the decrease this effect.}
\end{thebibliography}

%

\clearpage


\section{Electronic Physics Auxiliary Publication Service}

\section{The confined spherical $p$-spin Kac model} 

We present here the calculation of the action and of the saddle point equations of the spherical $p$-spin disordered model with Kac interactions~\cite{Kac} discussed in the paper. The model is constituted by $N$ soft spin variables $s_i\in \mathbb R$ located on the vertices of a cubic lattice. 
In the Kac model all the replica calculations, which are analogous to that of a standard $p$-spin model, 
are done keeping finite the interaction range, which is controlled by the length-scale $\gamma$. 
Then at the end one takes a saddle point approximation sending to infinity the range of interaction, $\gamma\rightarrow\infty$. 
The peculiarity of the Kac model is that also inhomogeneous solutions for the overlap field, exact in the limit $\gamma\rightarrow\infty$, can be found~\cite{KacSilvio}. 
The Hamiltonian of the model is

\begin{equation}
H=-\sum_{i_1<\dots<i_p}J_{i_1,\dots ,i_p}s_1\dots s_p
\end{equation}
where the $J_{i_1,\dots,i_p}$ are random variables extracted from a gaussian distribution with zero mean and variance $\sigma^2=\frac{p!}{2\gamma^{(p-1)d}}\sum_{k}\psi(\frac{k-i_1}{\gamma})\dots \psi(\frac{k-i_p}{\gamma})$ and $\psi(x)$ is a $3$d normalized gaussian function.  Indeed, even if the indices $i_1,\ldots,i_p$ runs over all the spins, only the couples of spins placed at sites $i$ and $j$ such that $|i-j|<\gamma$ effectively interact.
This model was introduced~\cite{KacSilvio} to study interfaces and nucleations in disordered systems using the well known Kac approximation technique originally exploited in the context of first-order transitions~\cite{firstorder}. The advantage of the Kac approximation is that one can study interfaces and inhomogeneous effects by solving the saddle point equations obtained imposing boundary conditions at a finite distance, for instance at the boundaries of a spherical cavity of finite radius $\ell$, and keeping a finite interaction range $\gamma$. The solutions are then exact in the limit where both the length-scales diverge, $\ell,\gamma\rightarrow \infty$ but with a fixed ratio $\hat{\ell}=\ell/\gamma$.\\

\subsection{Analysis of a spherical cavity for RB and AB conditions}
In order to analyze RB and AB conditions on equal footing we shall enforce that the configurations equilibrated inside the cavity at temperature $T=1/\beta$ are equal outside the cavity to an equilibrium configuration at temperature $T'=1/\beta'$. The RB and AB cases respectively correspond to $\beta'=0$ and $\beta'=\beta$.

The free energy of a spherical region, $B_{\ell}$, of radius $\ell$ at temperature $T=1/\beta$ in presence of a boundary arranged in an equilibrium configuration $\mathcal{C}'$ at temperature $T'=1/\beta'$ reads 
\begin{equation}
F(\ell,\beta,\beta')=-\frac{1}{\beta}\overline{\frac{1}{Z(\beta')}\sum_{\mathcal{C'}}\exp(-\beta' H[\mathcal{C}'])\log(Z_{\mathcal C'}(\beta))}
\end{equation}
where 
\begin{equation}
Z(\beta)=\sum_{\mathcal{C}}\exp(-\beta H[\mathcal C]) \ ,
\end{equation}
\begin{equation}
Z_{\mathcal{C}'}(\beta)=\sum_{\mathcal{C}}\exp(-\beta H[\mathcal C])\prod_{i\notin B(\ell)} \delta{s_i,s'_i} \ ,
\end{equation}
the overbar stands for the average operation over the choice of the couplings, and $\mathcal{C'}=\{s'_i\}$.
Two series of replicas are introduced to compute averages. According to standard replica manipulations (see~\cite{CGlong} for an analogous computation) the free energy reads
\begin{eqnarray}
F(\ell,\beta,\beta')=\hspace{6cm}\\ \nonumber-\lim_{n\rightarrow0}\frac{1}{\beta n} \overline{[\log Z^{n+1}_{\Omega}(\ell,\beta,\beta')-\log Z(\beta)]}=\\ \nonumber-\lim_{\substack{n\rightarrow0\\\nu\rightarrow0}}\frac{1}{\beta \nu n}\overline{[\left(Z_{\Omega}^{n+1}\right)^{\nu}-Z^{\nu}]}
\end{eqnarray}
with 
\begin{eqnarray}
Z_{\Omega}^{n+1}(\ell,\beta,\beta')=\hspace{6cm}\\ \nonumber\sum_{\mathcal C', \mathcal C^a}\exp\left(-\beta'H[\mathcal C']-\beta\sum_{a=1}^{n}H[\mathcal C^a]\right)\prod_{i\notin B(\ell)} \delta{s^a_i,s'_i} \ .
\end{eqnarray}
We now perform the Gaussian integration over the couplings, the continuum limit and 
assume replica symmetry over the $\nu$ replicas. This is justified by the fact that 
different $\nu$ replicas are subjected to different boundary conditions; hence they are not expected to clusters inside the same states. We can then introduce the collective overlap variables 
\begin{eqnarray}
q_{a,b}(x)d^dx=\frac{1}{\delta^d}\sum_{i\in B_x(\delta)}s_i^as_i^b\\
q'_a(x)d^dx=\frac{1}{\delta^d}\sum_{i\in B_x(\delta)}s_i^as'_i
\end{eqnarray}
where $B_x(\delta)$ is a cube of intermediate size $\delta$ such that $\gamma\ll\delta\ll\ell$.
Since we are only interested in relative values of the free-energy we drop all terms 
independent of $q_{a,b}$ and obtain
\begin{eqnarray}
\label{freenenrgy}
F(\ell,\beta,\beta')=\hspace{6.5cm}\\ \nonumber-\lim_{\substack{n\rightarrow0\\\nu\rightarrow0}}\frac{1}{\beta \nu n}\int \prod_{a<b}dq_{a,b}\exp(-\gamma^d\nu\mathcal{S}(q_{a,b};\ell,\beta,\beta'))
\end{eqnarray} 
with 
\begin{widetext}
\begin{eqnarray}
\mathcal{S}(q_{a,b};\ell,\beta,\beta')=\hspace{15cm}\\ \nonumber -\frac{1}{4}\int d^dx \left[\beta^2\sum_{a,b}\left(\int d^dyq_{a,b}(y)\psi(|x-y|)\right)^p+2\beta\beta'\sum_a\left(\int d^dyq'_a(y)\psi(|x-y|)\right)^p\right]-\frac{1}{2}\int d^dx\log\det[q_{a,b}(x)]
\label{action}
\end{eqnarray}
\end{widetext}
The free energy in~\eqref{freenenrgy} is dominated by the saddle point of the action $\mathcal{S}$ which gives the condition for the elements of the overlap matrix $q_{a,b}$.\\
We studied the solution of the problem in the two extremal cases with $\beta'=0$ (confinement case) and with $\beta'=\beta$ (amorphous boundary), which was already solved in~\cite{FranzMontanari}. While the equilibrium boundary case can be always solved with a replica symmetric \emph{ansatz},
in the confinement case, and in particular for small spherical cavities ($\ell<\ell_C$, see the paper), one needs a 1RSB {\it ansatz} for the overlap matrix: $q_{a,b}(x)=\delta_{ab}(1-q_1(x))+\delta_{\lambda(a)\lambda(b)}(q_1(x)-q_0(x))+q_0(x)$, where $\lambda(a)$ assigns different labels to the $n/m$ groups of $m$ replicas.
In the following subsection we present the explicit form of the action with the 1RSB parametrization and the corresponding saddle point equations 
for the overlap fields.

\subsection{Action and saddle point equations for the RB case}
Within the 1RSB {\it ansatz} and with $\beta'=0$ the action can be explicitly written as
\begin{widetext}
\begin{eqnarray}
\mathcal{S}_{1RSB}(q_1(x), q_0(x), m,n;\ell,\beta)=-\frac{n\beta^2}{2}\int d^dx\bigg[(m-1)f_x(q_1\ast\psi)+(n-m)f_x(q_0\ast\psi)\bigg]+\\ \nonumber-\frac{1}{2}\int d^d x\bigg[(m-1)\frac{n}{m}\log(1-q_1(x))+\left(\frac{n}{m}-1\right)\log(1-q_1(x)-m[q_1(x)-q_0(x)])+\bigg.\\ \nonumber \bigg.+\log(1-q_1(x)+m[q_1(x)-q_0(x)]+n q_0(x))\bigg]
\end{eqnarray}
\end{widetext}
where we defined $f_x(g)=g(x)^p/2$, and $q\ast\psi=\int d^d y q(y)\psi(|x-y|)$ is a function of $x$.\\
In the $n\rightarrow 0$ limit this action becomes proportional to $n$. Thanks to the large $\gamma$ prefactor the integral in~\eqref{freenenrgy} is evaluated in the saddle point approximation. Hence the free energy has to be evaluated in correspondence of the solutions of the equations that assure the stationarity of the action with respect to variations of $q_1$, $q_0$ and $m$. The final expression for the free energy in~\eqref{freenenrgy} and of the equations for $q_1(x)$, $q_0(x)$ and $m$ read as follows:
\begin{widetext}
\begin{eqnarray}
F(\ell,\beta)=-\frac{\beta}{2}\int d^dx\bigg[(m-1)f_x(q_1\ast\psi)-m f_x(q_0\ast\psi)\bigg]+\\ \nonumber-\frac{1}{2\beta}\int d^d x\bigg[\frac{m-1}{m}\log(1-q_1(x))+\frac{1}{m}\log(1-q_1(x)-m[q_1(x)-q_0(x)])+\frac{q_0(x)}{1-q_1(x)+m[q_1(x)-q_0(x)]}\bigg]
\end{eqnarray}
\begin{eqnarray}
\frac{\partial \mathcal{S}_{1RSB}|_{m=1}}{\partial q_1(x)}=0\rightarrow \beta^2 p \int d^dyf'_y(q_1\ast\psi)\psi(|x-y|)=\hspace{9cm}\\ \nonumber=\frac{1}{m}\left[\frac{1}{1-q_1(y)}-\frac{1}{1-q_0(y)+(m-1)(q_1(y)-q_0(y))}+\frac{m q_0(y)}{(1-q_0(y)+(m-1)(q_1(y)-q_0(y)))^2}\right]
\end{eqnarray}
\begin{eqnarray}
\frac{\partial \mathcal{S}_{1RSB}|_{m=1}}{\partial q_0(x)}=0\rightarrow \beta^2 p \int d^dyf'_y(q_0\ast\psi)\psi(|x-y|)=\frac{m q_0(y)}{(1-q_0(y)+(m-1)(q_1(y)-q_0(y)))^2}\hspace{3.4cm}
\end{eqnarray}
\begin{eqnarray}
\frac{\partial \mathcal{S}_{1RSB}|_{m=1}}{\partial m}=0\rightarrow \beta^2\int d^dx [f_x(q_1\ast\psi)-f_x(q_0\ast\psi)]+\frac{1}{m}\int d^d x \bigg[\frac{1}{m}\log\left(\frac{1-q_1(x)}{1-q_0(x)+(m-1)(q_1(x)-q_0(x))}\right)\bigg. \hspace{0.5cm}\\ \nonumber\bigg.+(q_1(x)-q_0(x))\left[\frac{1}{1-q_0(x)+(m-1)(q_1(x)-q_0(x))}-m\frac{q_0(x)}{(1-q_0(x)+(m-1)(q_1(x)-q_0(x)))^2}\right]\bigg]=0 \ .
\end{eqnarray}
\end{widetext}
We solved this equations in presence of the boundary condition $q_1(x)=q_0(x)=1$ everywhere outside a sphere of radius $\ell$ to ensure the restriction to the confinement case. In this case we also set $\beta'=0$ and we imposed that the overlap between the replica which characterizes the random boundary, and the replicas at temperature $\beta^{-1}$ is identically zero: $q'_a(x)=0 \ \ \ \forall x$.\\
The equilibrium boundary case already studied in~\cite{FranzMontanari} is obtained imposing $\beta'=\beta$. In this limit the RS {\it ansatz} is always stable and we imposed that $q'_a(x)=q_{a,b}(x)=q(x)$ with $a\in [1,n]$. Hence the action reads
\begin{widetext}
\begin{eqnarray}
\mathcal{S}(q(x), m;\ell,\beta)=-\frac{(m-1)\beta^2}{2}\int d^dx [ mf_x(q\ast\psi)]-\frac{m-1}{2}\int d^d x[\log(1-q(x))+q(x)] \ ,
\end{eqnarray}
\end{widetext}
and the free energy of the spherical region is 
\begin{eqnarray}
F(\ell,\beta)=\hspace{7cm}\\\nonumber-\frac{\beta}{2}\int d^dx [ mf_x(q\ast\psi)]-\frac{1}{2\beta}\int d^d x[\log(1-q(x))+q(x)]
\end{eqnarray}
with $q(x)$ such that $\partial \mathcal S|_{m-1}/\partial q(x)=0$ and hence 
\begin{eqnarray}
\beta^2 p \int d^dyf'_y(q\ast\psi)\psi(|x-y|)=\frac{q(y)}{1-q(y)} \ .
\end{eqnarray}

\section{Large $\ell_C$ limit} 
We now study the behavior of the solution in the limit of large $\ell_C$, {\it i.e } when $T\rightarrow T_K$. In this case, for $\ell>\ell_{C}$ the average overlap is equal to $q_0$. 
Its value in the center of the cavity therefore coincides with the bulk value, which is zero, 
up to exponentially small (in $\ell$) corrections due to the boundary condition. 
Below $\ell_C$, instead, the average overlap is equal to $(1-m)q_1+mq_0$. At the center of the cavity, $q_1$ is also equal to its bulk value, which is
the value obtained for the fully connected p-spin model, up to exponentially small (in $\ell$) corrections due to the boundary condition (this, as the previous statement, is what we expect and what is found by numerical integration). The breaking parameter $m$ instead decreases
slower with $\ell$ and thus determines the leading behavior in $\ell$ for the overlap 
at the center of the cavity for $\ell<\ell_C$. 
We shall now obtain its dependence on $\ell$.
In the limit of large $\ell_C$, {\it i.e } when $T\rightarrow T_K$, we can decompose the action in a volume term and a surface contribution: 
\begin{equation} 
\mathcal S(m^*,T) = - Y(m^*,T) \ell^{d-1} + B(m^*,T) \ell^d \ ,
\label{phenoaction}
\end{equation}
where the $m^*$ stands for the equilibrium value of m for each choice of $T$ and $\ell$ .\\
We can write two useful equations. The first means that the configurational entropy, obtained as the derivative in $m=1$, goes to zero at the ideal glass transition: 
\begin{equation} 
\frac{\partial B}{ \partial m}(m=1, T=T_K) = 0 \ .
\end{equation}
Note that the surface term does not plat any role in it, as it should since in this limit one consider infinite cavities and hence the bulk behavior is recovered.
The second provides the definition of $\ell_C$: 
\begin{equation} 
- \frac{\partial Y}{ \partial m} + \frac{\partial B}{ \partial m} \ell_{C} = 0 \rightarrow \ell_{C} =  \frac{\partial Y}{ \partial m}(1,T) / \frac{\partial B}{ \partial m}(1,T) \ .       
\label{pointtoset}
\end{equation} 
\begin{figure}
\hspace{-1.0cm}
\includegraphics[scale=0.33,angle=-90]{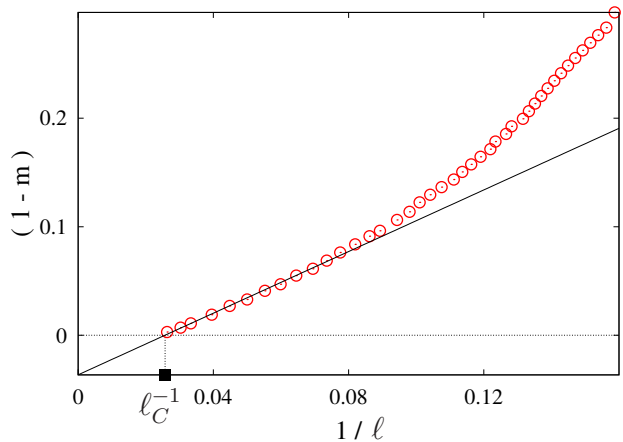}
\put(-190,-158){\large $\ell_C^{-1}$}
\put(-102,-167){\large $\ell$}
\caption{\label{fig3} Symmetry breaking parameter $(1-m)$ as a function of the inverse cavity radius $1/\ell$ at $T=0.594$, $\ell_C=38.85$.}
\vspace{-0.7cm}
\end{figure}
For $\ell<\ell_{C}$, $m^*$ (that we shall simply call $m$ from now on) is such that $\partial \mathcal S/\partial m=0$. \\ 
To study how $m$ moves from $1$ as soon as $\ell\lesssim\ell_{C}$ and for very large $\ell_{C}$ ({\it i.e.} very small $T-T_K$), we expand this relation around $m=1$ and $T=T_K$:
\begin{eqnarray} 
- \frac{\partial Y}{ \partial m} - \frac{\partial^2 Y}{ \partial m^2}(m-1) - \frac{\partial^2 Y}{ \partial m \partial T} (T-T_k) 
+ \hspace{2cm}\\\nonumber +\ell \left[ \frac{\partial B}{ \partial m}+\frac{\partial^2 B}{ \partial m^2}(m-1)  + \frac{\partial^2 B}{ \partial m \partial T} (T-T_k)   \right]+... = 0
\end{eqnarray}  
where all the derivatives have been computed in $m=1$ and $T=T_k$ and hence $\partial B/\partial m=0$.
The resulting equation reads:
\begin{eqnarray} 
\label{approx}
(m-1) \left[ - \frac{\partial^2 Y}{ \partial m^2} + \ell \frac{\partial^2 B}{ \partial m^2} \right] =\hspace{4cm}\\\nonumber \frac{\partial Y}{ \partial m} - \frac{\partial^2 B}{ \partial m \partial T} (T-T_k) \ell +\frac{\partial^2 Y}{ \partial m \partial T} (T-T_k) \ .  
\end{eqnarray}
Also expanding around $T=T_K$, where $\partial B/\partial m=0$, the denominator of~\eqref{pointtoset} we obtain at the leading order
\begin{equation}
\ell_{C} =\dfrac{ \frac{\partial Y}{ \partial m} }{ \frac{\partial^2 B}{ \partial m \partial T}} (T-T_k)^{-1}\ .
\end{equation}
This gives for~\eqref{approx} the following result:
\begin{equation} 
(m-1) \left[ - \frac{\partial^2 Y}{ \partial m^2} + \ell \frac{\partial^2 B}{ \partial m^2} \right] = \frac{\partial Y}{ \partial m} \left(1- \frac{\ell}{\ell_{C}}\right) + O(T-T_K) \ .
\end{equation}
Hence in the large $\ell_{C}$ (small $T-T_K$) limit the previous equation reduces to 
\begin{equation} 
m \simeq 1+\dfrac{\frac{\partial Y}{ \partial m} }{\frac{\partial^2 B}{ \partial m^2} }\frac{1}{\ell}\left(1- \frac{\ell}{\ell_{C}}\right) \ .
\end{equation}
In particular, for large radii close to the confinement length, $\ell \gg 1$ and $\ell \lesssim \ell_C$, we expect from the above equation that 
\begin{equation}
 1 - m \sim  \frac{1}{\ell}-\frac{1}{\ell_{C}},
\end{equation} 
a behaviour that can be clearly seen in the Fig.1. 
\section{An heuristic analysis of the relaxation time-scale inside the cavity}
We now sketch some heuristic arguments to understand the evolution of the relaxation timescale for the confined system as a function of $\ell$. We will focus on the richer regime
corresponding to temperatures below $T_h$ and above $T_{MCT}$.
As already discussed in the text, we expect that by decreasing $\ell$ the system first undergoes a MCT cross-over and then starts to show a truly activated dynamics. In the latter regime one can repeat the usual RFOT theory arguments to obtain the relaxation time as a function of $\ell_{PS}$ but now one has to consider the point-to-set length for the confined 
system, {\it i.e.} for RB conditions. This will be different (larger) than the one of the bulk system. We neglect inhomogeneous effects to simplify the discussion and simply consider that the configurational entropy density for the confined system reads $s_c-3Y_C/\ell$. 
By repeating the usual arguments leading to the point-to-set length-scale one obtains 
$$
\ell_{PS}=\frac{3Y_{PS}}{s_c-3Y_C/\ell}
$$
where again for simplicity we assume $\theta_C=2$. This equation shows that indeed 
$\ell_{PS}$ for the confined system increases by decreasing $\ell$. Assuming the usual 
(dynamic scaling) exponential relation between time and length-scale one directly finds that
the relaxation time-scale also increases by decreasing $\ell$. As we have stressed before,
no real transition can happen in a finite size system so the divergence of $\ell_{PS}$ and, hence, of the relaxation time should be cut-off at a certain point for a finite range of the interaction $\gamma$ (In the strict Kac-limit this instead is not the case). In a real system (no infinite interaction range) we heuristically expect that the cut-off takes place when $\ell$ reaches the value $\ell_{PS}$, thus leading to the result
$$
\ell_{cut-off}=\frac{3Y_{PS}+3Y_C}{s_c}
$$
Since numerically we find that $Y_{PS}\simeq Y_C$ the previous equations implies that when 
the size of the confined region reaches twice the point to set the growth of the relaxation time
should be cut-off. This argument certainly is very crude and needs to be refined but gives a first crude idea of the behavior of the relaxation time as a function of $\ell$.

\end{document}